\documentclass[12pt, a4paper]{article}

\usepackage[english]{babel}
\usepackage{lineno}
\usepackage{graphicx}
\usepackage{mathrsfs,dsfont,mathtools,bm,braket,amsmath,amsfonts}
\usepackage[dvipsnames]{xcolor}
\usepackage[babel,kerning=true,spacing=true]{microtype}
\usepackage{physics}
\usepackage{siunitx}
\usepackage[top=2.5cm,bottom=2.5cm,left=2.5cm,right=2.5cm]{geometry}
\usepackage[colorlinks=true, allcolors=blue]{hyperref}
\usepackage{float}
\usepackage{authblk} 

\title{\textbf{Mode Conversion of Gaussian Beams at Dielectric Interfaces}}

\author[1]{Eli Meril}
\affil[1]{\small \textit{School of Physics and Astronomy, Tel Aviv University, Tel Aviv 69978, Israel}}
\date{\vspace{-5ex}} % Removes the date for a cleaner IOP look

\begin{document}
\maketitle
\begin{abstract}
We investigate mode conversion of $\mathrm{TEM}_{00}$ Gaussian beams upon transmission through planar dielectric interfaces. We show that the angle-dependent Fresnel coefficients act as a spatial filter, inevitably generating higher-order spatial modes. Using a vector angular spectrum formulation and numerical simulations, we reveal that this polarization-dependent filtering induces a coupling from $\mathrm{TEM}_{00}$ into higher-order Laguerre-Gaussian modes, yielding a quadrupolar field pattern. We quantify the associated amplitude and phase deviations, showing that the mode fidelity decreases significantly as the beam waist approaches the diffraction limit.
\end{abstract}

\vspace{1em}

\section{Introduction}
In many areas of precision optical physics, from nanoscale metrology to high-resolution confocal microscopy, a Gaussian beam incident on a planar dielectric interface is typically treated by applying plane-wave Fresnel coefficients to the entire beam \cite{Vardanyan2011, AbuSafia1994, Eismann2025}. Within this approximation, the reflected and transmitted fields remain in the fundamental $\mathrm{TEM}_{00}$ mode. Although mathematically and numerically convenient, this assumption neglects essential physics. A finite-waist beam is not a single plane wave but a coherent superposition of plane waves spanning a continuum of propagation directions. Each component encounters the interface at a different angle of incidence and is weighted by a different Fresnel coefficient. Consequently, this surface functions as an angle-dependent filter. Because this filtering applies non-uniformly across the field spectrum, it directly couples the $\mathrm{TEM}_{00}$ mode into higher-order modes. As a result, the angular spectrum of the beam is modified during transmission or reflection. For weakly focused beams, this effect is negligible; however, it becomes significant for tightly focused beams, such as those encountered in high-numerical-aperture optical systems \cite{Zonnevylle2013}, or when the spectrum approaches regions where the Fresnel coefficients vary rapidly, for example near the Brewster or critical angles.

\vspace{6pt}

In this work, we develop an analytical Vector Angular Spectrum (VAS) formulation to quantify this mode conversion. We show that angle-dependent Fresnel coefficients generate higher-order spatial modes, producing amplitude and phase deviations in the transmitted field. Numerical simulations confirm these predictions, revealing a characteristic quadrupolar spatial signature arising directly from the polarization-dependent angular filtering of the interface.

\section{Vectorial Interface Filtering} \label{sec:vas}
Consider an $x$-polarized monochromatic Gaussian beam with a waist $w_0$ (at $z=0$) and wavenumber $k_1 = n_1 \omega / c$, incident on a planar dielectric interface. Within the VAS framework \cite{hohenester2020nano, RichardsWolf1959}, the electric field is represented as a continuum of plane waves. The angular spectrum amplitude of the fundamental $\mathrm{TEM}_{00}$ mode is given by
\begin{equation}
    \tilde{E}_{i}(k_x, k_y, z=0) =  E_0 \frac{w_0^2}{2} \exp\left(-\frac{w_0^2 k_\perp^2}{4}\right),
    \label{E_k_space}
\end{equation}
where $k_\perp = \sqrt{k_x^2 + k_y^2}$ denotes the transverse wavevector magnitude. 
The transversality condition, $\nabla\!\cdot\!\mathbf{E}=0$, constrains the field by requiring that $\tilde{E}_z=-\frac{k_x\tilde{E}_x+k_y\tilde{E}_y}{k_z}, 
k_z=\sqrt{k_1^2-k_\perp^2}.
$ Each spectral component corresponds to a plane wave incident at an angle $\theta_i = \arcsin\!\left(\frac{k_\perp}{k_1}\right)$.

To evaluate the reflected and transmitted fields, each incident plane wave is projected onto the local $s$ (TE) and $p$ (TM) polarization basis associated with its specific wave vector. The transmitted field is then constructed as the coherent sum of these components, appropriately weighted by their respective angle-dependent Fresnel transmission coefficients, $t_s(\theta_i)$ and $t_p(\theta_i)$:
\begin{equation}
    \tilde{\bm{E}}^t_i(k_x, k_y) = t_p(\theta_i) \tilde{\bm{E}}_{i}^p + t_s(\theta_i) \tilde{\bm{E}}_{i}^s.
    \label{trans_E}
\end{equation}
Since both $t_p(\theta_i) $ and $ t_s(\theta_i)$ vary with the incidence angle $\theta_i$, the interface inherently acts as an angle-dependent spatial filter, causing the transmitted spectrum to deviate from its original Gaussian profile. An analogous filtering occurs for the reflected field $\tilde{\bm{E}}^r_i$ upon replacing the transmission coefficients with $r_p(\theta_i)$ and $r_s(\theta_i)$.

\section{Analytical Model}
\subsection{Scalar Fresnel Filtering}
To quantify the mode conversion induced by transmission through a planar dielectric interface, we first consider a scalar model of a normally incident $\mathrm{TEM}_{00}$ Gaussian beam. Our goal is to obtain an analytic expression for the transmitted field inside the second medium. Vectorial corrections are incorporated in the following subsection.

In the paraxial limit, the Fresnel transmission coefficient can be expanded as
\begin{equation}
    t(k_\perp) \approx t_0 + \beta k_\perp^2,
    \label{Fresnel_transmission_coefficients}
\end{equation}
where $t_0=t(0)$ is the standard normal-incidence transmission coefficient and $\beta=(2k_1^2)^{-1}\partial_\theta^2 t(0)$.
Multiplication by $k_\perp^2$ in momentum space corresponds, via the Fourier derivative theorem, to applying the transverse Laplacian in real space. Using Eq.~(\ref{trans_E}), the transmitted field therefore becomes
$$
E_t(r) \approx t_0E_i(r) - \beta\nabla_\perp^2 E_i(r).
$$
For an incident Gaussian beam, $\nabla_\perp^2 E_i(r)  = \frac{4}{w_0^2} \left( \frac{r^2}{w_0^2} - 1 \right) E_i(r)$, so that the transmitted field acquires higher-order spatial modes,
\begin{equation}
    E_t(r) \approx \underbrace{\left[ t(0) + \frac{1}{(k_1 w_0)^2} \left. \frac{\partial^2 t}{\partial \theta_i^2} \right|_{0} \right] E_i(r)}_{\text{Modified Fundamental TEM}_{00}} + \underbrace{\left[ \frac{1}{(k_1 w_0)^2} \left. \frac{\partial^2 t}{\partial \theta_i^2} \right|_{0} \right] L_1^0\left(\frac{2r^2}{w_0^2}\right) E_i(r)}_{\text{Coupled LG}^0_{1} \text{ Mode}}
\end{equation}
Thus, within the scalar approximation, the interface couples the fundamental $\mathrm{TEM}_{00}$ mode to the radial ${LG}^0_{1}$ mode. The coupling amplitude scales as $(k_1 w_0)^{-2}$ and therefore vanishes in the plane-wave limit.

\subsection{Vectorial Correction and Quadrupolar Structure}
The scalar treatment captures only the rotationally symmetric component of the Fresnel filter. However, as shown in Fig.\ref{fig:transmission_analysis}, vectorial simulations reveal an additional four-lobe spatial structure. This symmetry breaking is a direct consequence of the azimuthal dependence of the local $s$ and $p$ polarization bases associated with each transverse wavevector in momentum space. To account for this polarization-induced asymmetry, we extend the previous scalar treatment to a rigorous vectorial description.

Let $\mathbf k_\perp = k_\perp(\cos\phi_k,\sin\phi_k)$ and define the local polarization unit vectors as
\begin{equation}
\hat{\mathbf s}=(-\sin\phi_k,\cos\phi_k,0),
\qquad
\hat{\mathbf p}_i=(\cos\theta_i\cos\phi_k,\cos\theta_i\sin\phi_k,-\sin\theta_i).
\end{equation}
For a normally incident beam polarized along $\hat{\mathbf x}$, projecting the transmitted field onto Cartesian coordinates yields
\begin{equation}
\tilde E_{t,x}
\simeq
\tilde E_0(k_\perp)
\left[
t_p(\theta_i)\cos^2\phi_k+t_s(\theta_i)\sin^2\phi_k
\right],
\label{eq:Etx_phi}
\end{equation}
\begin{equation}
\tilde E_{t,y}
\simeq
\tilde E_0(k_\perp)
\left[
t_p(\theta_i)-t_s(\theta_i)
\right]
\sin\phi_k\cos\phi_k.
\label{eq:Ety_phi}
\end{equation}
Expanding the Fresnel coefficients near normal incidence,
\begin{equation}
t_p(k_\perp)\approx t_0+\beta_p k_\perp^2,
\qquad
t_s(k_\perp)\approx t_0+\beta_s k_\perp^2.
\end{equation}
Explicit evaluation (see the appendix for full dervation) gives
\[
\beta_s = \frac{1}{k_1^2}\frac{n_1(n_1-n_2)}{n_2(n_1+n_2)},
\qquad
\beta_p = \frac{n_1}{n_2}\beta_s .
\]
Using $k_\perp^2\cos 2\phi_k = k_x^2-k_y^2$ and $ k_\perp^2\sin 2\phi_k = 2k_xk_y,$ the transmitted field becomes
\begin{equation}
\tilde E_{t,x}
=
\tilde E_0
\left[
t_0+\frac{\beta_p+\beta_s}{2}k_\perp^2
+\frac{\beta_p-\beta_s}{2}(k_x^2-k_y^2)
\right],
\end{equation}
\begin{equation}
\tilde E_{t,y}
=
\tilde E_0
(\beta_p-\beta_s)k_xk_y.
\end{equation}
Transforming these expressions back to real space yields
\begin{equation}
E_{t,x}
=
t_0E_i
-\frac{\beta_p+\beta_s}{2}\nabla_\perp^2E_i
-\frac{\beta_p-\beta_s}{2}(\partial_x^2-\partial_y^2)E_i,
\label{eq:Ex_real_vector}
\end{equation}
\begin{equation}
E_{t,y}
=
-(\beta_p-\beta_s)\partial_x\partial_y E_i.
\label{eq:Ey_real_vector}
\end{equation}
Applying these derivatives to the incident $\mathrm{TEM}_{00}$ mode yields
\begin{equation}
E_{t,x}^{\mathrm{aniso}}\propto (x^2-y^2)e^{-r^2/w_0^2}
\propto r^2\cos 2\varphi\,e^{-r^2/w_0^2},
\end{equation}
\begin{equation}
E_{t,y}^{\mathrm{aniso}}\propto xy\,e^{-r^2/w_0^2}
\propto r^2\sin 2\varphi\,e^{-r^2/w_0^2},
\end{equation}
where $\varphi$ is the azimuthal angle in real space. 
Combining these derivations, the full transmitted field components are:
\begin{equation}
\begin{aligned}
E_{t,x}(r,\varphi) \approx\;&
\left[
t_0+\frac{1}{2(k_1 w_0)^2}
\left.
\left(
\frac{\partial^2 t_p}{\partial \theta_i^2}
+
\frac{\partial^2 t_s}{\partial \theta_i^2}
\right)
\right|_{0}
\right] E_i(r)
\\
&+
\left[
\frac{1}{2(k_1 w_0)^2}
\left.
\left(
\frac{\partial^2 t_p}{\partial \theta_i^2}
+
\frac{\partial^2 t_s}{\partial \theta_i^2}
\right)
\right|_{0}
\right]
L_1^0\!\left(\frac{2r^2}{w_0^2}\right) E_i(r)
\\
&-
\left[
\frac{1}{(k_1 w_0)^2}
\left.
\left(
\frac{\partial^2 t_p}{\partial \theta_i^2}
-
\frac{\partial^2 t_s}{\partial \theta_i^2}
\right)
\right|_{0}
\right]
\frac{r^2}{w_0^2}\cos 2\varphi \, E_i(r),
\\[4pt]
E_{t,y}(r,\varphi) \approx\;&
-
\left[
\frac{1}{(k_1 w_0)^2}
\left.
\left(
\frac{\partial^2 t_p}{\partial \theta_i^2}
-
\frac{\partial^2 t_s}{\partial \theta_i^2}
\right)
\right|_{0}
\right]
\frac{r^2}{w_0^2}\sin 2\varphi \, E_i(r),
\end{aligned}
\label{eq:full_vector_result}
\end{equation}
The first two terms describe the rotationally symmetric ($m=0$) contribution featuring the $LG_0^1$ mode, while the remaining terms correspond to a quadrupolar contribution with azimuthal order $m=\pm2$. These $LG_0^{\pm2}$ components are responsible for the four-lobed residual pattern observed in our numerical simulations. The total magnitude of this mode conversion is controlled by the parameter $(k_1 w_0)^{-2}$, while the anisotropic component scales with the polarization asymmetry $\beta_p-\beta_s$.

\section{Quantifying Mode Deviation}
To quantify the deviation of the beam from the fundamental $\text{TEM}_{00}$ mode, we calculate the mode overlap integral, or mode fidelity $\eta$, between the transmitted field $\mathbf{E}_t$ and a reference Gaussian field $\bm{E}$:
\begin{equation}
    \eta = \frac{\left| \iint \bm{E}^* \cdot \bm{E}_t \, dA \right|^2}{\left( \iint |\bm{E}|^2 \, dA \right) \left( \iint |\bm{E}_t|^2 \, dA \right)}.
    \label{fidelity}
\end{equation}
The parameter $\eta$ represents the power fraction coupled to the $\text{TEM}_{00}$ mode. It serves as a rigorous metric to evaluate coupling efficiencies in single-mode optical systems such as fibers \cite{Fan2020} or resonant cavities \cite{Gehr2010}.

\section{Numerical Results}
To validate our analytical predictions, we performed vectorial angular spectrum simulations of a normally incident Gaussian beam transmitted through a planar dielectric interface. The incident beam is an $x$-polarized $\mathrm{TEM}_{00}$ mode with wavelength $\lambda=532\,\mathrm{nm}$, propagating in air ($n_1=1$) toward a silicon interface with complex refractive index $4.12 + 0.048\mathrm{i}$.
\vspace{6pt}

\noindent
At the interface plane, the incident field is decomposed into its angular spectrum. Each spectral component is then transmitted using the exact Fresnel coefficients, according to the vector angular spectrum formulation described in Sec.~\ref{sec:vas}.

\vspace{6pt}

\noindent We consider tightly focused beams approaching the diffraction limit. In particular, we choose a beam waist $w_0=\lambda/2\approx0.266\,\mu\mathrm{m}$, representative of focusing conditions in high-numerical-aperture optical systems such as confocal microscopy, optical trapping, and near-field microscopy, where beams are routinely focused to near-wavelength scales.

\vspace{6pt}

\begin{figure}[htbp]
    \centering
    \includegraphics[width=1.05\linewidth]{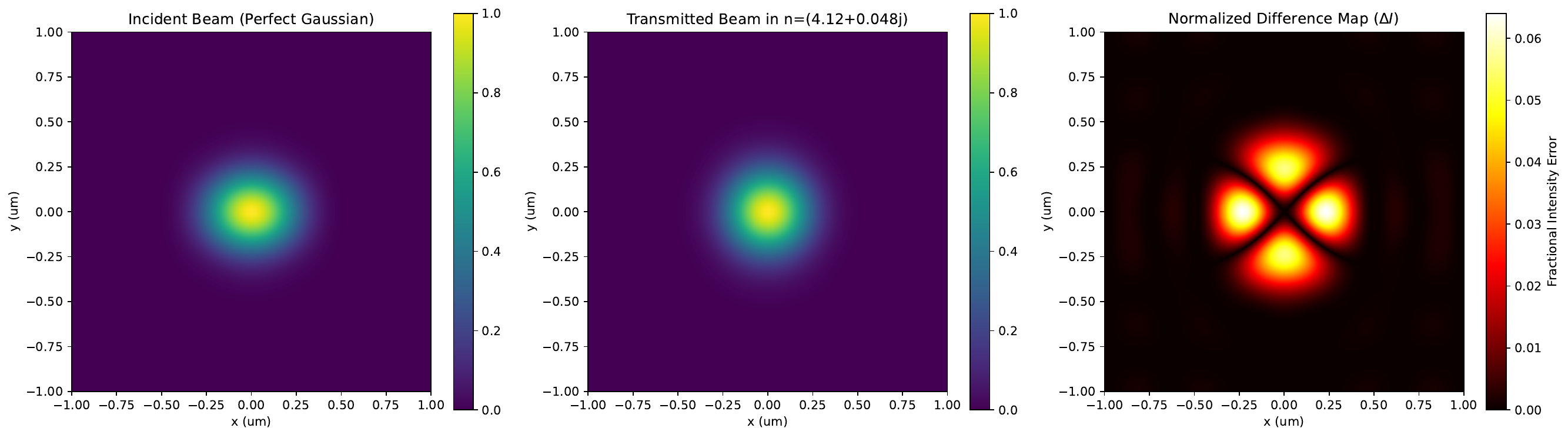}
    \caption{Transmitted beam profile for a tightly focused incident Gaussian beam with waist $w_0=\lambda/2$. Left: incident $\mathrm{TEM}_{00}$ mode intensity. Center: normalized transmitted intensity. Right: Intensity difference between the transmitted and the incident beam.}
    \label{fig:transmission_analysis}
\end{figure}

\noindent Figure~\ref{fig:transmission_analysis} shows the transmitted intensity distribution at the interface. Instead of preserving a pure $\mathrm{TEM}_{00}$ profile, the beam exhibits a quadrupolar deviation, indicating a coupling of the fundamental mode to higher-order spatial modes.

\begin{figure}[H]
    \centering
    \includegraphics[width=0.69\linewidth]{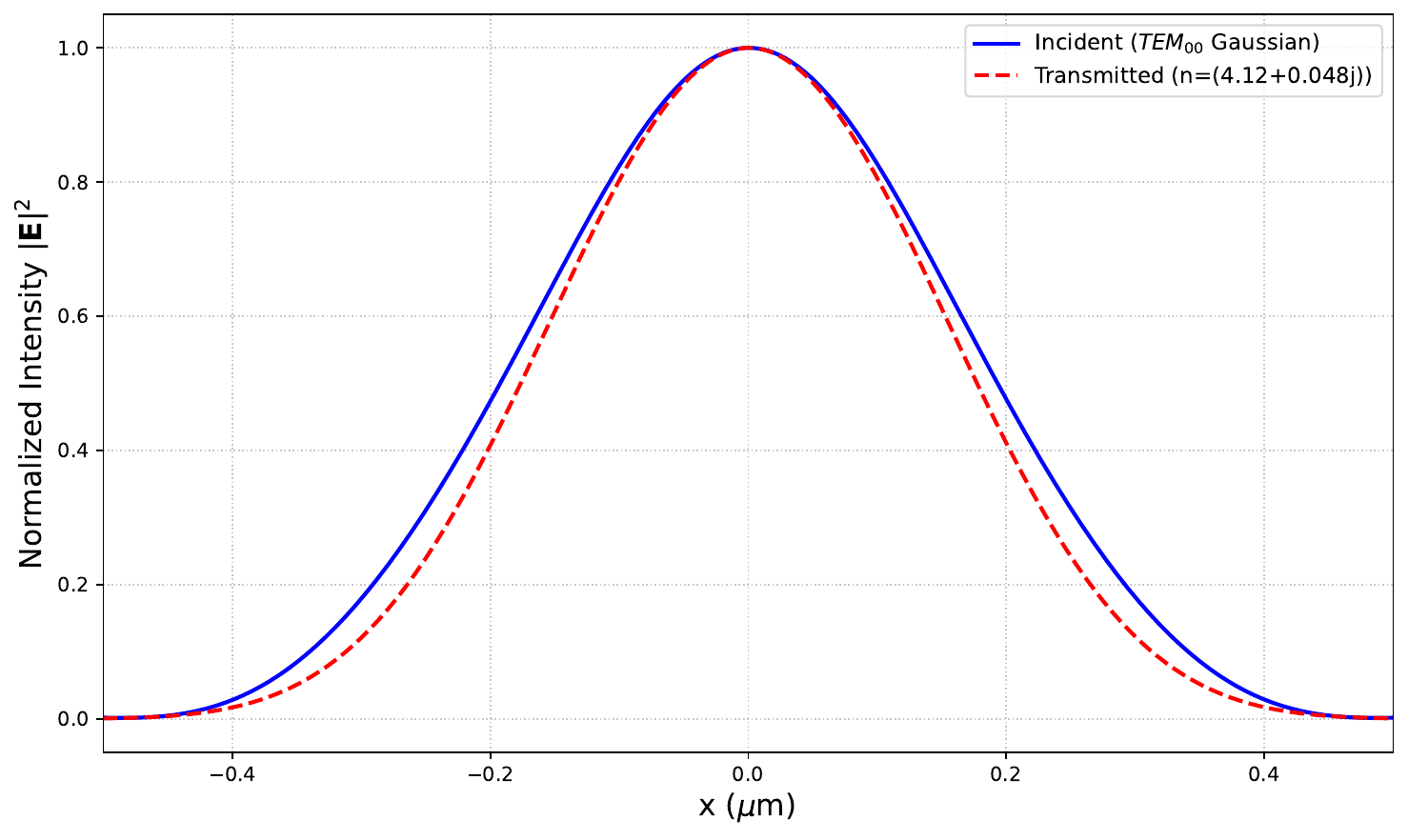}
    \caption{Normalized intensity profile of the incident and transmitted beams along the $x$-axis.}
    \label{fig:transmission_vs_x}
\end{figure}

\noindent The spatial deviation is further illustrated in one-dimensional intensity cut along the $x$-axis (Fig.~\ref{fig:transmission_vs_x}), where the transmitted beam clearly deviates from the incident Gaussian profile.

\begin{figure}[H] 
    \centering
    \includegraphics[width=0.73\linewidth]{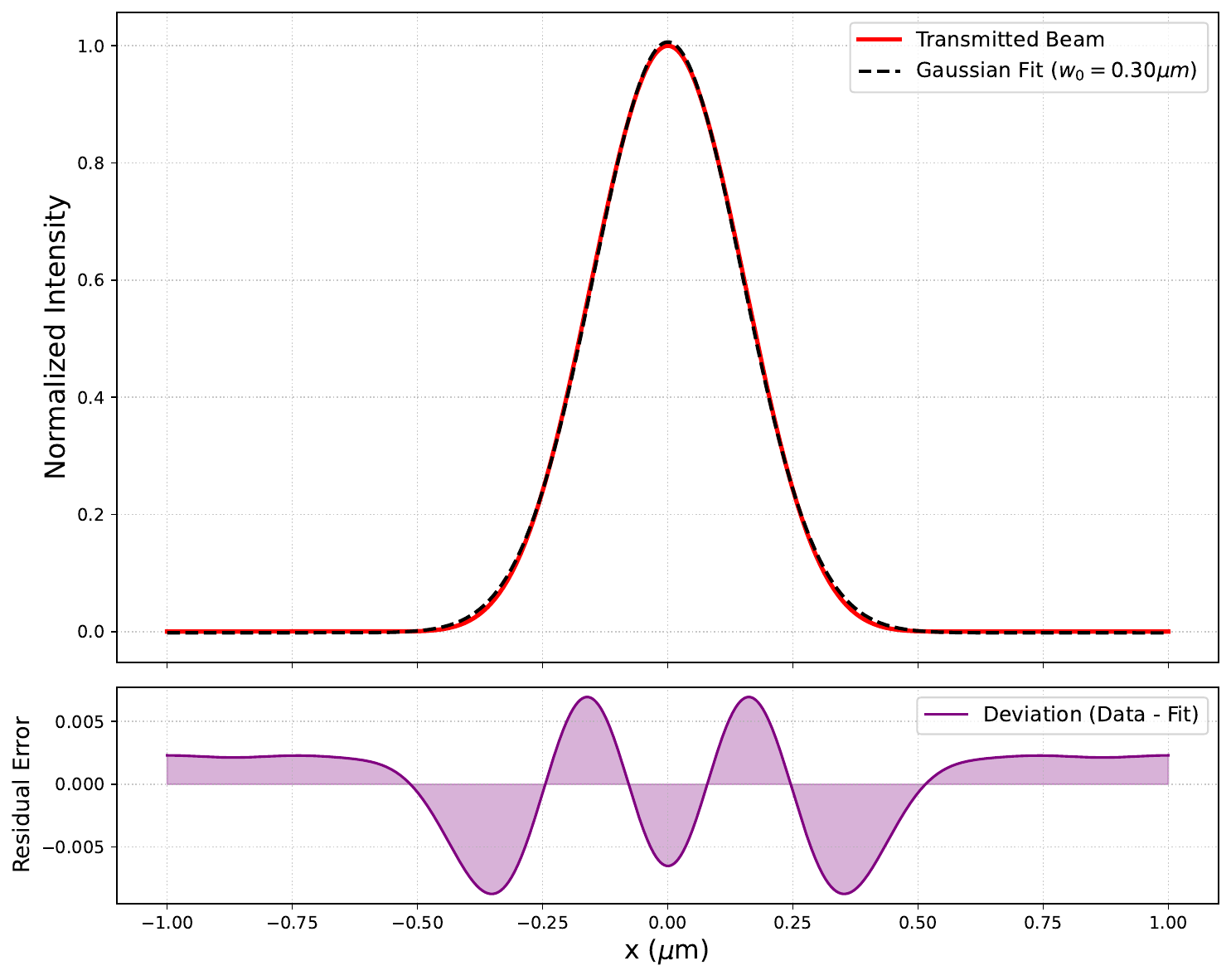}
    \caption{Transmitted beam intensity fitted to an ideal Gaussian profile. The lower panel shows the residual map, highlighting the non-Gaussian spatial structure.}
    \label{fig:transmission_fitted}
\end{figure}

\begin{figure}[H] 
    \centering
    \includegraphics[width=0.58\linewidth]{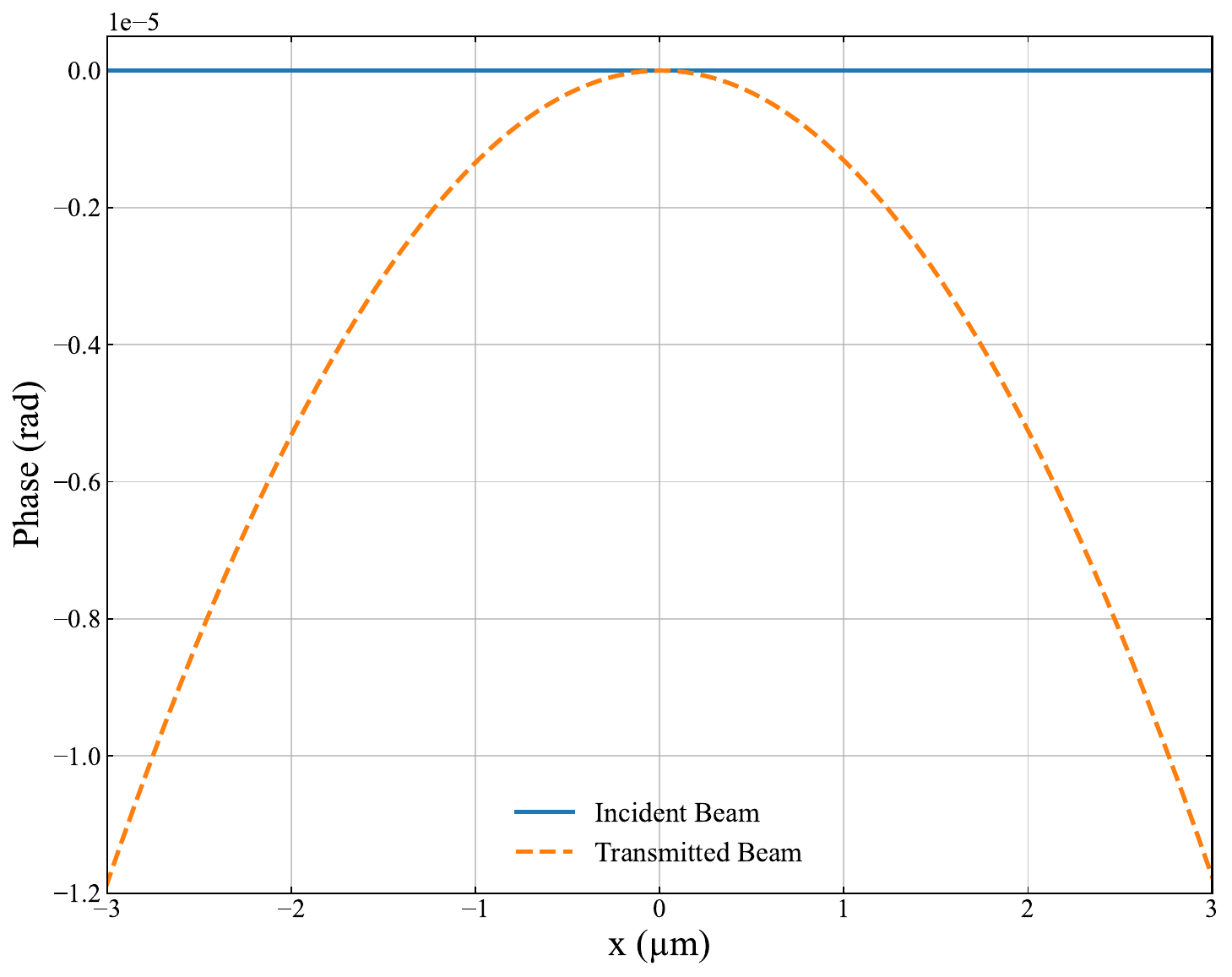}
    \caption{Phase profile of the transmitted beam along the $x$-axis at the interface ($z=0$). The incident beam has a flat phase front at the waist, whereas the transmitted field acquires a spatially varying phase due to the complex Fresnel coefficients.}
    \label{fig:phase_x_axis}
\end{figure}

\begin{figure}[H] 
    \centering
    \includegraphics[width=0.69\linewidth]{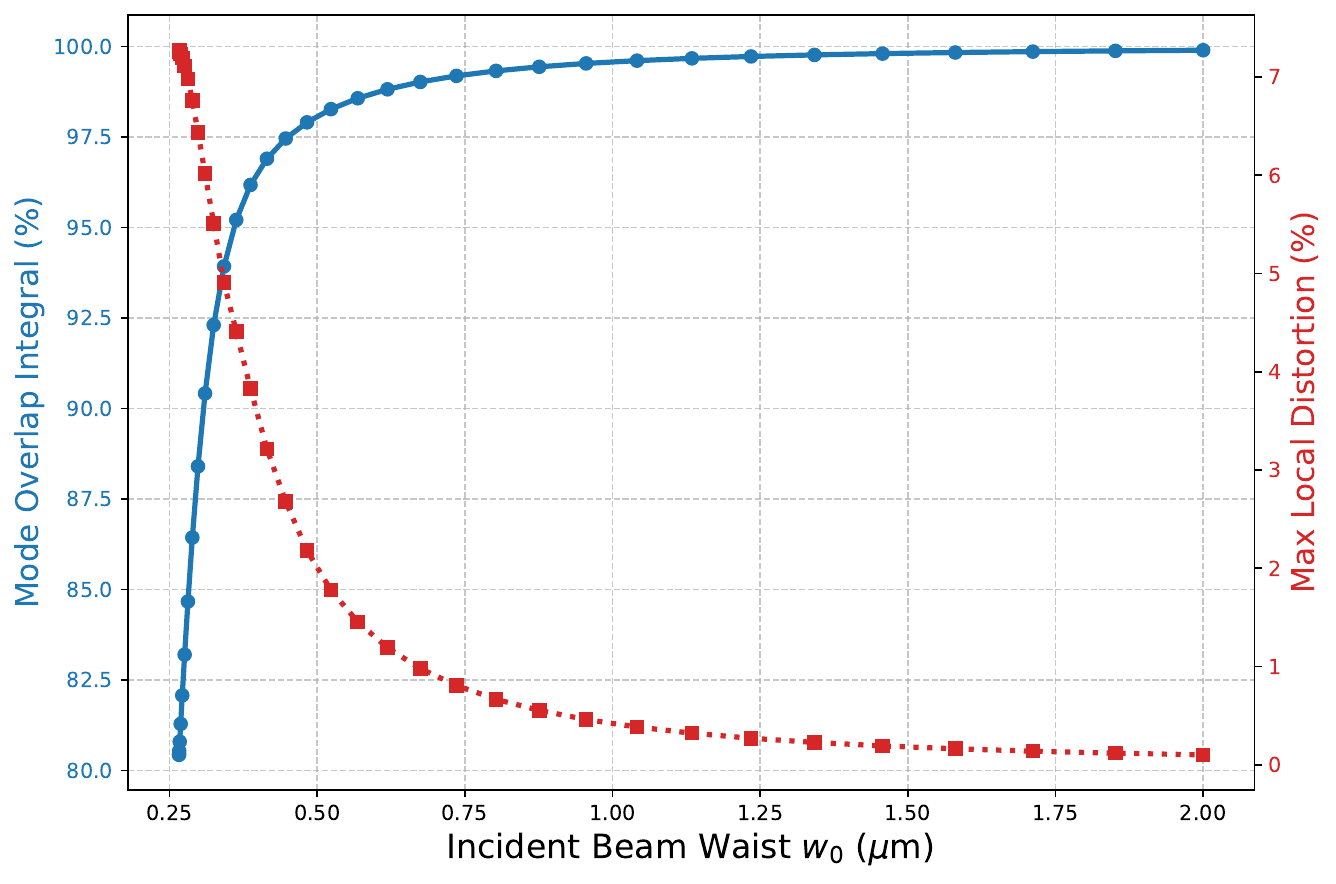}
    \caption{Mode overlap (fidelity) between the transmitted field and the incident field as a function of the incident beam waist $w_0$.}
    \label{fig:overlap_vs_waist}
\end{figure}

\noindent To quantify this deviation from a Gaussian mode, we fit the transmitted intensity to an ideal $\mathrm{TEM}_{00}$ profile. The resulting residual map (Fig.~\ref{fig:transmission_fitted}) exhibits systematic non-Gaussian structure, confirming the presence of higher-order spatial components.
\noindent In addition to amplitude distortions, the interface also induces phase variations due to the complex Fresnel transmission coefficients. As shown in Fig.~\ref{fig:phase_x_axis}, the transmitted field acquires a spatially varying phase even though the incident beam has a flat phase front at its waist.
\noindent Finally, we quantify the deviation from a pure Gaussian mode using the overlap fidelity defined in Eq.~(\ref{fidelity}). Figure~\ref{fig:overlap_vs_waist} shows the fidelity between the transmitted field and the ideal incident $\mathrm{TEM}_{00}$ mode as a function of the incident beam waist. For large waists ($k_1 w_0 \gg 1$), the fidelity approaches unity. However, as the beam approaches the wavelength scale, the fidelity decreases significantly, reflecting the increasing importance of nonparaxial mode coupling induced by the interface.
\noindent These numerical results confirm the analytical prediction that a dielectric interface acts as a momentum-dependent spatial filter, inevitably generating higher-order spatial modes when tightly focused beams interact with the boundary.

\section{Conclusion}
We have demonstrated that transmission through a planar dielectric interface inherently induces mode conversion in a fundamental $\mathrm{TEM}_{00}$ Gaussian beam. Using a vector angular spectrum formulation supported by numerical simulations, we show that the angular dependence of the Fresnel transmission coefficients acts as a momentum-dependent spatial filter. For tightly focused beams, this filtering drives the coupling of the fundamental $\mathrm{TEM}_{00}$ mode into higher-order spatial modes, leading to amplitude and phase deviations. Contrary to the standard paraxial assumption, this polarization-dependent filtering explicitly breaks the $SO(2)$ rotational symmetry.

\vspace{6pt}

These effects vanish in the plane-wave limit but become significant when the beam waist approaches the wavelength scale. Our results therefore reveal a source of spatial aberration at dielectric interfaces that is often neglected in paraxial modeling. Importantly, this mechanism arises even in ordinary homogeneous, isotropic, non-magnetic materials and does not rely on engineered photonic structures. These findings highlight the need to account for interface-induced mode coupling in high–numerical-aperture optical systems and precision photonic applications.

\newpage
\bibliographystyle{iopart-num}
\bibliography{sample}
\end{document}

\maketitle

% Initialize the supplementary counters
\beginsupplement

\section{Analytical Model and Derivations}
\label{appendix:A}
Consider the electric field of a fundamental $\mathrm{TEM}_{00}$ Gaussian beam evaluated at its waist ($z=0$):
\begin{equation*}
\bm{E}_i(r) = \hat{x} E_0 \exp\left(-\frac{r^2}{w_0^2}\right)
\end{equation*}
where $r^2 = x^2 + y^2$ and $w_0$ is the beam waist radius. 

Decomposing this spatial profile via a two-dimensional Fourier transform yields its angular spectrum of plane waves. In momentum space, where $k_\perp = \sqrt{k_x^2 + k_y^2}$ represents the transverse wavevector magnitude, the incident amplitude spectrum takes the form:
\begin{equation}
    \tilde{E}_i(k_\perp) = E_0 \frac{w_0^2}{2} \exp\left(-\frac{w_0^2 k_\perp^2}{4}\right).
\end{equation}
Upon interaction with a planar dielectric interface, each plane-wave component undergoes transmission governed by an angle-dependent Fresnel coefficient, $t(\theta_i)$. The angle of incidence $\theta_i$ is mapped to the transverse momentum via the exact geometric relation $\sin\theta_i = k_\perp / k_1$, where $k_1 = n_1 \omega / c$ is the wavenumber in the incident medium.

For a paraxial beam, the angular spectrum is tightly confined near normal incidence ($k_\perp \ll k_1$). Consequently, the Fresnel transmission coefficient can be expanded in a Taylor series around $\theta_i = 0$. Because an isotropic dielectric interface possesses rotational symmetry with respect to the surface normal, the transmission coefficient must be an even function of the incidence angle, $t(\theta_i) = t(-\theta_i)$. As a result, the first derivative and all higher odd-order derivatives vanish.

Applying the paraxial approximation $\sin(\theta_i) \approx \theta_i \approx k_\perp / k_1$ and truncating at second order, the transmission filter in momentum space becomes
\begin{equation}
    t(k_\perp) \approx t(0) + \frac{1}{2 k_1^2} \left. \frac{\partial^2 t}{\partial \theta_i^2} \right|_{0} k_\perp^2 \equiv t_0 + \beta k_\perp^2,
\end{equation}
where $t_0 \equiv t(0)$ is the standard normal-incidence transmission coefficient, and $\beta \equiv (2 k_1^2)^{-1} \partial_{\theta}^2 t(0)$.
The transmitted field in $k$-space is the product of the incident spectrum and the transmission filter:
$$\tilde{E}_t(k_\perp) = t(k_\perp) \tilde{E}_i(k_\perp) \approx \left( t_0 + \beta k_\perp^2 \right) \tilde{E}_i(k_\perp)$$
A common approximation applies the Fresnel coefficients directly to the Cartesian field components, which neglects the local polarization basis associated with each plane-wave component.
To recover the observable real-space field, we apply the inverse Fourier transform. By the Fourier derivative theorem
\begin{equation}
    E_t(r) \approx t_0 E_i(r) - \beta \nabla_\perp^2 E_i(r).
    \label{E_t(r)}
\end{equation}
Evaluating the transverse Laplacian for the fundamental Gaussian field yields:
$$\nabla_\perp^2 E_i(r)  = \frac{4}{w_0^2} \left( \frac{r^2}{w_0^2} - 1 \right) E_i(r)$$
To physically interpret this spatial distortion, we project the correction onto the orthogonal Laguerre-Gaussian ($\text{LG}_{p}^l$) mode basis. The radial polynomial for the lowest-order azimuthally symmetric mode is $L_1^0(x) = 1 - x$. For our Gaussian envelope with scaled argument $x = 2r^2/w_0^2$, the polynomial mapping is exactly:
\begin{equation}
    \frac{r^2}{w_0^2} - 1 = -\frac{1}{2} \left[ L_1^0\left(\frac{2r^2}{w_0^2}\right) + 1 \right].
\end{equation}
Substituting this relation into the Laplacian expression and recalling the definition of $\beta$, we arrive at the leading-order analytical expression for the transmitted field:
$$\nabla_\perp^2 E_i(r) = -\frac{2}{w_0^2} \left[ L_1^0\left(\frac{2r^2}{w_0^2}\right) + 1 \right] E_i(r)$$
Substituting in Eq.\ref{E_t(r)}, we arrive at the leading-order analytical expression for the transmitted field:
$$E_t(r) \approx t_0 E_i(r) + \frac{2\beta}{w_0^2} \left[ L_1^0\left(\frac{2r^2}{w_0^2}\right) + 1 \right] E_i(r)$$
$$E_t(r) \approx \underbrace{\left[ t(0) + \frac{1}{(k_1 w_0)^2} \left. \frac{\partial^2 t}{\partial \theta_i^2} \right|_{0} \right] E_i(r)}_{\text{Modified Fundamental TEM}_{00}} + \underbrace{\left[ \frac{1}{(k_1 w_0)^2} \left. \frac{\partial^2 t}{\partial \theta_i^2} \right|_{0} \right] L_1^0\left(\frac{2r^2}{w_0^2}\right) E_i(r)}_{\text{Coupled LG}^0_{1} \text{ Mode}}$$
\subsection{Vector-Induced Angular Correction}
\label{appendix:A_vector}
Let the transverse wavevector be written in polar coordinates as $\mathbf{k}_\perp = k_\perp(\cos\phi_k,\sin\phi_k)$. For each plane-wave component, the local polarization unit vectors are:
\begin{align}
    \hat{\mathbf s} &= (-\sin\phi_k, \cos\phi_k, 0), \\
    \hat{\mathbf p}_i &= (\cos\theta_i\cos\phi_k, \cos\theta_i\sin\phi_k, -\sin\theta_i).
\end{align}
For an incident beam polarized along $\hat{\mathbf x}$ where $\tilde{\mathbf E}_i(\mathbf{k}_\perp) = \tilde E_0(k_\perp)\hat{\mathbf x}$, the projections onto the local polarization basis are:
\begin{equation}
    \hat{\mathbf x}\cdot\hat{\mathbf s} = -\sin\phi_k, \qquad \hat{\mathbf x}\cdot\hat{\mathbf p}_i = \cos\theta_i\cos\phi_k.
\end{equation}
Retaining the leading transverse terms near normal incidence, the transmitted field is constructed as:
\begin{equation}
    \tilde{\mathbf E}_t(\mathbf{k}_\perp) \simeq \tilde E_0(k_\perp) \left[ -t_s(\theta_i)\sin\phi_k\,\hat{\mathbf s} + t_p(\theta_i)\cos\phi_k\,\hat{\mathbf p}_t \right].
\end{equation}
Projecting this back onto the transverse Cartesian components yields:
\begin{align}
    \tilde E_{t,x} &\simeq \tilde E_0(k_\perp) \left[ t_p(\theta_i)\cos^2\phi_k + t_s(\theta_i)\sin^2\phi_k \right], \label{eq:Etx_phi} \\
    \tilde E_{t,y} &\simeq \tilde E_0(k_\perp) \left[ t_p(\theta_i) - t_s(\theta_i) \right] \sin\phi_k\cos\phi_k. \label{eq:Ety_phi}
\end{align}
Introducing the symmetric and antisymmetric transmission combinations, $t_\pm(\theta_i) = \frac{1}{2}[t_p(\theta_i) \pm t_s(\theta_i)]$, Eqs.~(\ref{eq:Etx_phi}) and (\ref{eq:Ety_phi}) can be elegantly rewritten as:
\begin{align}
    \tilde E_{t,x} &= \tilde E_0(k_\perp) \left[ t_+(\theta_i) + t_-(\theta_i)\cos 2\phi_k \right], \\
    \tilde E_{t,y} &= \tilde E_0(k_\perp) \, t_-(\theta_i)\sin 2\phi_k.
\end{align}
To evaluate $\beta_s$, we begin with the exact Fresnel transmission coefficient for $s$-polarization. Eliminating the angle of refraction via Snell's law, the coefficient is expressed entirely in terms of the incident angle $\theta_i$:
\begin{equation}
    t_s(\theta_i) = \frac{2 n_1 \cos\theta_i}{n_1 \cos\theta_i + n_2 \sqrt{1 - \left(\frac{n_1}{n_2}\right)^2 \sin^2\theta_i}}.
\end{equation}
Applying the small-angle expansions yields:
\begin{equation}
    t_s(\theta_i) \approx \frac{2 n_1 \left(1 - \frac{1}{2}\theta_i^2 \right)}{n_1 \left(1 - \frac{1}{2}\theta_i^2\right) + n_2 \left[1 - \frac{1}{2}\left(\frac{n_1}{n_2}\right)^2 \theta_i^2\right]}.
\end{equation}
Expanding this to leading order in $\theta_i^2$:
\begin{equation}
    t_s(\theta_i) \approx t_s(0) \left[ 1 + \frac{1}{2}\theta_i^2 \left( \frac{n_1}{n_2} - 1 \right) \right].
\end{equation}
Mapping the incidence angle back to the transverse momentum ($\theta_i \approx k_\perp / k_1$), the transmission filter takes the explicit form:
\begin{equation}
    t_s(k_\perp) \approx t_s(0) + \left[ \frac{1}{k_1^2} \frac{n_1(n_1 - n_2)}{n_2(n_1 + n_2)} \right] k_\perp^2.
\end{equation}
A parallel derivation for p-polarization provides the corresponding parameters:
\begin{equation}
    \beta_s = \frac{1}{k_1^2} \frac{n_1(n_1 - n_2)}{n_2(n_1 + n_2)}, \qquad \beta_p = \frac{n_1}{n_2} \beta_s.
\end{equation}
Near normal incidence ($t_p(0)=t_s(0)=t_0$), substituting the parabolic expansions $t_{p,s}(k_\perp) \approx t_0 + \beta_{p,s} k_\perp^2$ gives:
\begin{equation}
    t_+(k_\perp) \approx t_0 + \frac{\beta_p+\beta_s}{2}k_\perp^2, \qquad t_-(k_\perp) \approx \frac{\beta_p-\beta_s}{2}k_\perp^2.
\end{equation}
Utilizing the Cartesian mappings $k_\perp^2\cos 2\phi_k = k_x^2-k_y^2$ and $k_\perp^2\sin 2\phi_k = 2k_xk_y$, the leading-order transmitted field in $k$-space becomes:
\begin{align}
    \tilde E_{t,x} &= \tilde E_0 \left[ t_0 + \frac{\beta_p+\beta_s}{2}k_\perp^2 + \frac{\beta_p-\beta_s}{2}(k_x^2-k_y^2) \right], \\
    \tilde E_{t,y} &= \tilde E_0 (\beta_p-\beta_s)k_xk_y.
\end{align}
Transforming back to real space replaces the momentum variables with spatial derivatives:
\begin{align}
    E_{t,x} &= t_0E_i - \frac{\beta_p+\beta_s}{2}\nabla_\perp^2E_i - \frac{\beta_p-\beta_s}{2}(\partial_x^2-\partial_y^2)E_i, \label{eq:Ex_real_vector} \\
    E_{t,y} &= -(\beta_p-\beta_s)\partial_x\partial_y E_i. \label{eq:Ey_real_vector}
\end{align}
The first correction in Eq.~(\ref{eq:Ex_real_vector}) is the rotationally symmetric term derived in the scalar model. The subsequent terms are inherently anisotropic and vanish only when the interface exhibits no polarization asymmetry ($\beta_p=\beta_s$).

Applying these anisotropic operators to the Gaussian input field $E_i(x,y)=E_0\exp[-(x^2+y^2)/w_0^2]$ generates:
\begin{equation}
    (\partial_x^2-\partial_y^2)E_i = \frac{4(x^2-y^2)}{w_0^4}E_i, \qquad \partial_x\partial_y E_i = \frac{4xy}{w_0^4}E_i.
\end{equation}
Converted to polar coordinates $(r, \varphi)$, the real-space vector corrections take the final form:
\begin{align}
    E_{t,x}^{\mathrm{aniso}} &\propto (x^2-y^2)e^{-r^2/w_0^2} \propto r^2\cos 2\varphi \, e^{-r^2/w_0^2}, \\
    E_{t,y}^{\mathrm{aniso}} &\propto xy\,e^{-r^2/w_0^2} \propto r^2\sin 2\varphi \, e^{-r^2/w_0^2}.
\end{align}
This rigorous quadrupolar structure is the precise origin of the four-lobe residual pattern observed in the numerical simulations.

\subsection{Effective Beam Width Approximation}
An alternative perspective can be derived by treating the transmission coefficient as a momentum-space Gaussian filter. Factoring the normal-incidence transmission $t_0$ from the truncated expansion yields:
$$t(k_\perp) = t_0 \left( 1 + \frac{\beta}{t_0} k_\perp^2 \right)$$
For a tightly localized paraxial spectrum, the perturbation term $\frac{\beta}{t_0} k_\perp^2$ is small, allowing us to apply the approximation $1 + x \approx \exp(x)$. The transmission filter is thus elegantly approximated as:
$$t(k_\perp) \approx t_0 \exp\left( \frac{\beta}{t_0} k_\perp^2 \right)$$
Multiplying the incident angular spectrum by this effective transmission filter provides the transmitted field in $k$-space:
\begin{align*}
\tilde{E}_t(k_\perp) &= t(k_\perp) \tilde{E}_i(k_\perp) \\
&\approx \left[ t_0 \exp\left( \frac{\beta}{t_0} k_\perp^2 \right) \right] \left[ E_0 \frac{w_0^2}{2} \exp\left(-\frac{w_0^2}{4} k_\perp^2\right) \right] \\
&\approx t_0 E_0 \frac{w_0^2}{2} \exp\left[ -\frac{w_0^2}{4} k_\perp^2 + \frac{\beta}{t_0} k_\perp^2 \right]
\end{align*}
By inspecting the Gaussian argument, we can immediately identify an effective transmitted beam waist, $w_t$:
$$w_t = w_0 \sqrt{1 - \frac{4\beta}{t_0 w_0^2}}$$
We define the dimensionless Fresnel aberration number
\begin{equation}
C_F = \frac{\beta}{t_0 w_0^2},
\end{equation}
which quantifies the strength of interface-induced spatial filtering. From the form of the effective waist, it is evident that this approximation remains valid only in the weak-aberration limit, $C_F \ll 1$.

\section{Additional Numerical Analysis}
\label{appendix:B}
This appendix provides supplementary numerical data quantifying the maximum deviation of the transmitted field from an ideal incident Gaussian profile across varying beam parameters and propagation conditions.
\begin{figure}[htbp]
    \centering
    \includegraphics[width=0.56\linewidth]{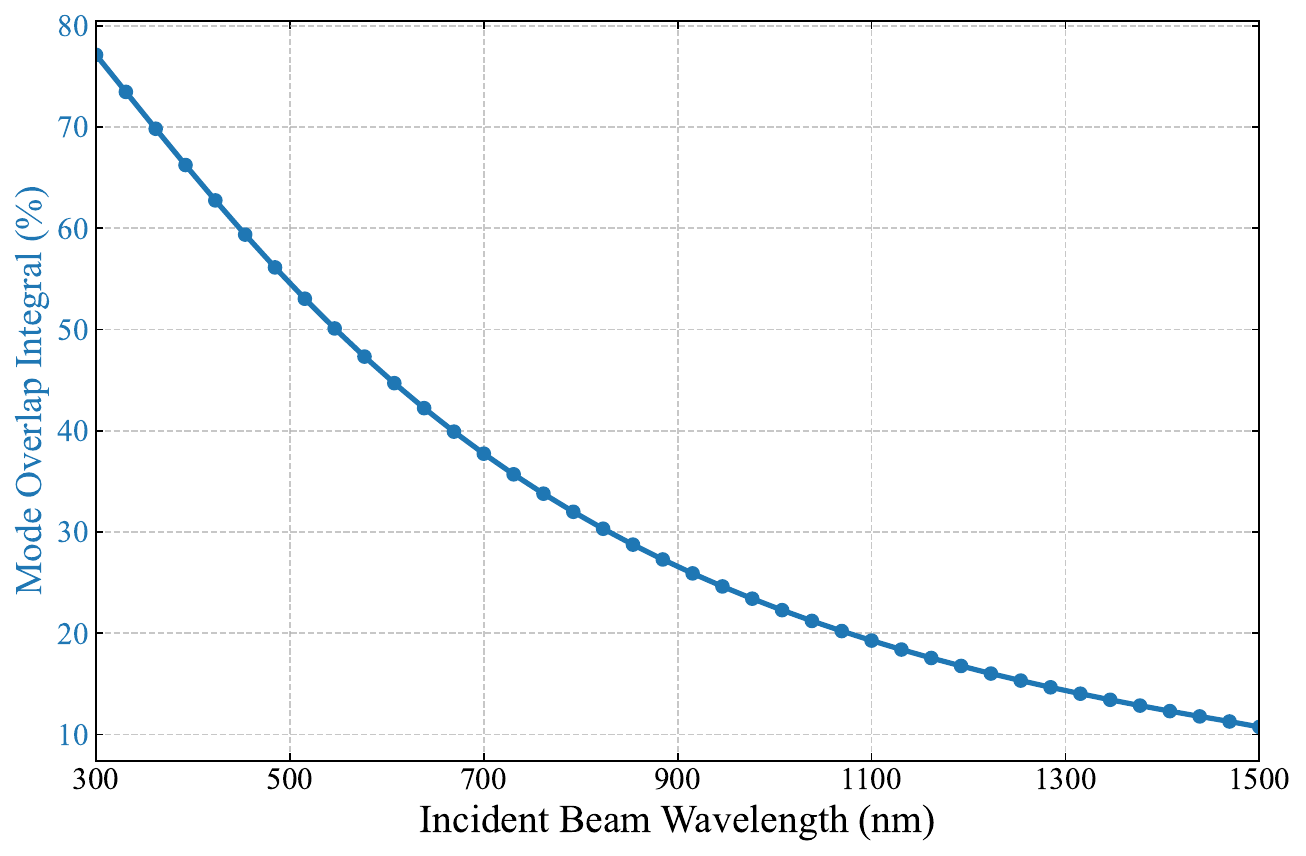}
    \caption{Mode overlap fidelity as a function of the incident beam wavelength $\lambda$.}
    \label{fig:fidelity_vs_wavelength}
\end{figure}
We now quantify the deviation between the incident and transmitted beams as a function of propagation distance within the second medium.
\begin{figure}[htbp]
    \centering
    \includegraphics[width=0.59\linewidth]{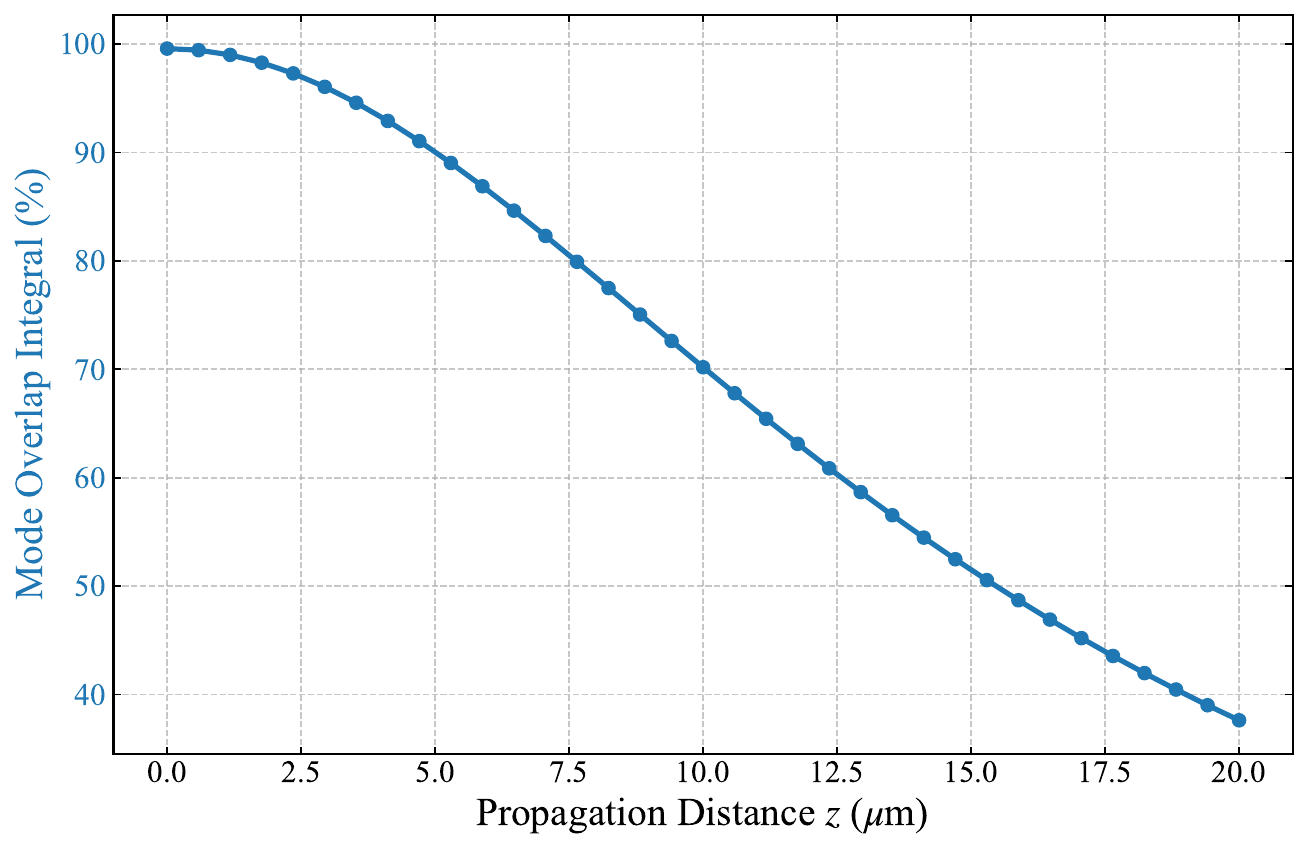}
    \caption{Mode overlap fidelity of the transmitted beam as a function of propagation distance $z$ within the silicon medium.}
    \label{fig:fidelity_vs_z}
\end{figure}
Finally, we map the overarching parameter space comparing the incident and transmitted beam fidelity.
\begin{figure}[htbp]
    \centering
    \includegraphics[width=0.8\linewidth]{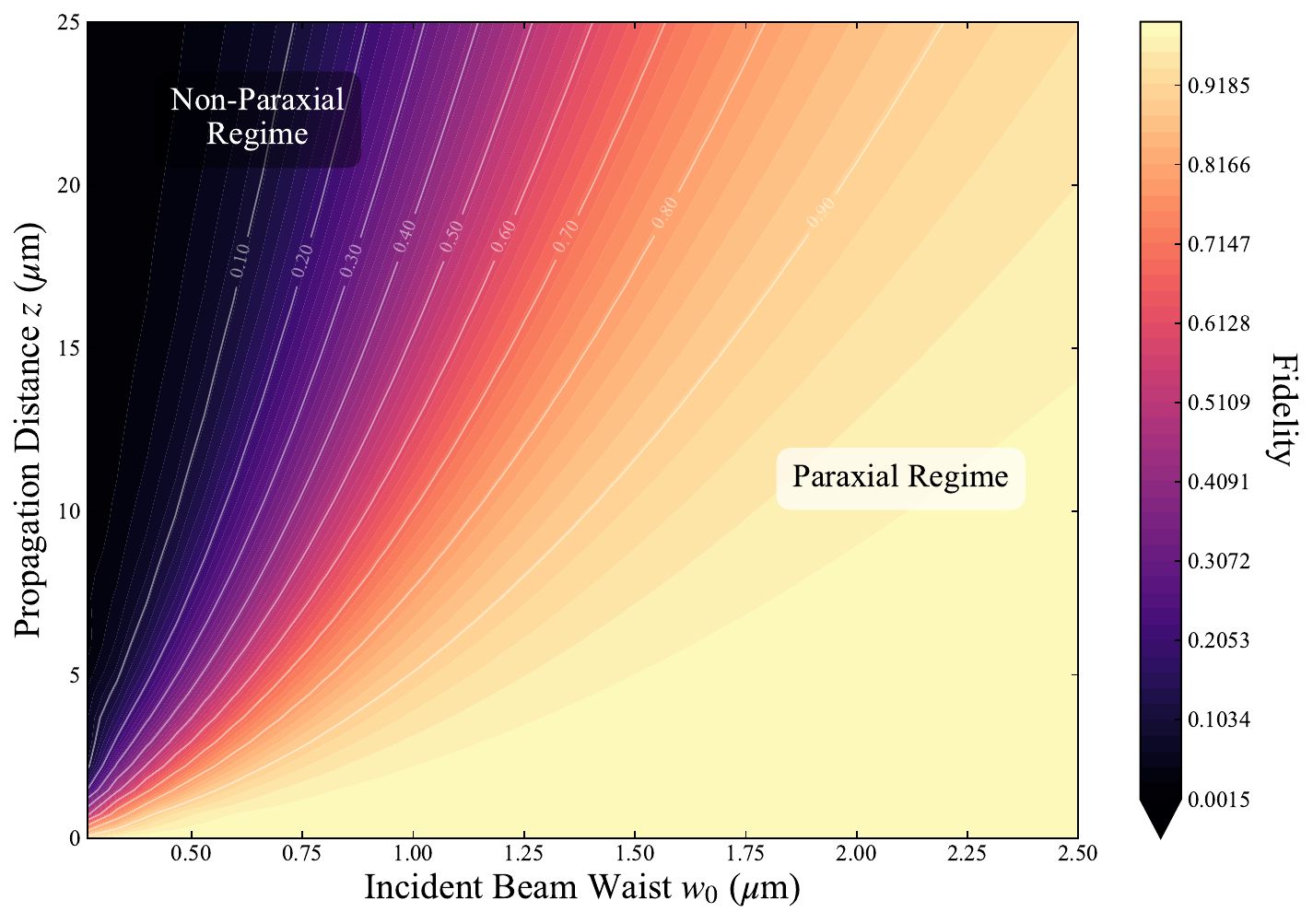}
    \caption{Two-dimensional parameter space mapping of the transmission fidelity between the ideal incident beam and the resulting transmitted field.}
    \label{fig:fidelity_2d_map}
\end{figure}